\documentclass[sigconf,screen,nonacm]{acmart}

\usepackage{booktabs}
\usepackage{tikz}
\usepackage{pgfplots}
\pgfplotsset{compat=1.16}
\usetikzlibrary{arrows.meta, positioning, shapes.geometric, calc, backgrounds, decorations.pathreplacing}
\usepackage{listings}
\usepackage{xcolor}
\usepackage{url}
\usepackage{enumitem}
\hypersetup{hidelinks}

\begin{document}

\title{Codified Context: Infrastructure for AI Agents in a Complex Codebase}

\author{Aristidis Vasilopoulos}
\affiliation{%
  \institution{Independent Researcher}
  \country{USA}
}
\email{arisvas4@gmail.com}

\begin{abstract}
LLM-based agentic coding assistants lack persistent memory: they lose coherence across sessions, forget project conventions, and repeat known mistakes. Recent studies characterize how developers configure agents through manifest files, but an open challenge remains how to \emph{scale} such configurations for large, multi-agent projects. This paper presents a three-component \emph{codified context infrastructure} developed during construction of a 108,000-line C\# distributed system: (1)~a hot-memory constitution encoding conventions, retrieval hooks, and orchestration protocols; (2)~19 specialized domain-expert agents; and (3)~a cold-memory knowledge base of 34 on-demand specification documents. Quantitative metrics on infrastructure growth and interaction patterns across 283 development sessions are reported alongside four observational case studies illustrating how codified context propagates across sessions to prevent failures and maintain consistency. The framework is published as an open-source companion repository.
\end{abstract}


\maketitle

\section{Introduction}
\label{sec:introduction}

AI coding agents such as GitHub Copilot, Cursor, and Claude Code~\cite{anthropic2026claudecode} have reached millions of developers, and recent work documents fully agentic systems capable of planning, executing, and iterating on complex development tasks~\cite{dong2025codegen, robbes2026agentic}. These tools possess broad programming knowledge, but they lack project memory: each session begins without awareness of prior sessions, established conventions, or past mistakes. Consistent output for a specific project requires knowledge that persists across sessions, yet single-file manifests (\texttt{.cursorrules}, \texttt{CLAUDE.md}, \texttt{AGENTS.md}) do not scale beyond modest codebases~\cite{santos2025decoding, chatlatanagulchai2025manifests, jiang2025cursor, chatlatanagulchai2025readmes}: a 1,000-line prototype can be fully described in a single prompt, but a 100,000-line system cannot. The AI must be told---repeatedly, reliably, and in a format it can act on---how the project works, what patterns to follow, and what mistakes to avoid. Structured knowledge transfer to agents remains a largely open interaction design problem~\cite{huang2025professional}. This paper addresses the gap with a codified context infrastructure that treats documentation as \emph{infrastructure}---load-bearing artifacts that AI agents depend on to produce correct output. Machine-readable specification documents, available on demand, allow agents to simulate persistent memory even in a complex codebase.

The architecture was developed iteratively during construction of a 108,000-line C\# distributed system (a real-time multiplayer simulation built on the MonoGame framework and the Arch Entity Component System library). Both application code and context infrastructure were generated using Claude Code as the sole code-generation tool, directed by human prompting and agent orchestration. The author's primary background is in chemistry rather than software engineering, making this project a test case for a specific emerging use pattern: domain experts building software beyond their primary expertise with AI agents.

\subsection{Contributions}
\label{sec:contributions}

\begin{enumerate}
\item \textbf{A tiered architecture for organizing project knowledge to support multi-agent AI-assisted development.} This architecture extends the single-file manifest pattern with domain-expert agents embedding project-specific knowledge, trigger tables for automatic task routing, and a hot/cold memory separation that distinguishes always-loaded conventions from on-demand specifications.
\item \textbf{Quantitative evaluation across 283 development sessions}, including infrastructure growth metrics, interaction patterns (2,801 human prompts, 1,197 agent invocations, 16,522 agent turns), and four observational case studies.
\item \textbf{An open-source framework} with representative agent specifications, an MCP retrieval server, example documents, factory agents for bootstrapping, and all analysis scripts.
\end{enumerate}

\section{Related Work}
\label{sec:related}

\subsection{Agentic Coding Manifests}
\label{sec:manifests}

Developers have begun creating configuration files---variously called \texttt{CLAUDE.md}, \texttt{.cursorrules}, or \texttt{AGENTS.md}---to provide AI coding agents with project-specific instructions at the start of each session. Several empirical studies now characterize these files. Among Claude Code projects, 72.6\% specify application architecture~\cite{santos2025decoding}, and the pattern generalizes across 2,303 files spanning Claude Code, Codex, and GitHub Copilot~\cite{chatlatanagulchai2025manifests, chatlatanagulchai2025readmes}. A classification of the types of instructions developers include has been developed from 401 Cursor repositories~\cite{jiang2025cursor}. Adoption across the broader open-source ecosystem remains early---only ${\sim}$5\% of 466 surveyed repositories had adopted any context file format~\cite{mohsenimofidi2025context}---but among projects that do use manifests, quantitative evidence of effectiveness is emerging: the presence of AGENTS.md files was associated with a 29\% reduction in median runtime and 17\% reduction in output token consumption~\cite{lulla2026agentsmd}.

These studies characterize what developers write in manifest files. The present work addresses a different question: what happens when a project's knowledge needs outgrow a single file? The project described here began with a manifest similar to those analyzed in~\cite{santos2025decoding, chatlatanagulchai2025manifests} but evolved into a tiered architecture totaling approximately 26,000 lines---more than an order of magnitude beyond the typical manifests characterized in prior studies. The recently released Google Conductor~\cite{google2026conductor} for the Gemini CLI addresses a similar problem through persistent Markdown and a structured spec-plan-implement workflow. The present work, developed independently and concurrently, focuses on a tiered knowledge organization designed to be portable across agentic coding tools rather than coupled to a specific platform.

\subsection{Context Engineering, Multi-Agent Frameworks, and LLM-Assisted Software Engineering}
\label{sec:context-engineering}

Integrated multi-tool workflows for context engineering in multi-file code generation have shown higher success rates than single-agent systems~\cite{haseeb2025context}. Augmenting LLMs with codified human expert domain knowledge improves output quality as well~\cite{ulanuulu2026augmenting}. The approach taken here aligns with the principle that codified domain knowledge improves agent output~\cite{ulanuulu2026augmenting} but operates at project scale: the knowledge base typically captures project-specific conventions, architectural decisions, and known failure modes rather than general domain expertise. Context engineering has been formalized as a discipline with a taxonomy of foundational components drawn from over 1,400 papers~\cite{mei2025context}. Agentic Context Engineering (ACE) treats contexts as ``evolving playbooks'' refined through a generate-reflect-curate cycle~\cite{zhang2026ace}. That work also identifies a \emph{brevity bias}---a tendency for iterative optimization to collapse toward short, generic prompts---which is consistent with the finding here that specialized agents require substantial embedded domain knowledge to perform reliably (Section~\ref{sec:agents}).

Multi-agent coordination frameworks such as AutoGen~\cite{wu2023autogen}, ChatDev~\cite{qian2023chatdev}, and MetaGPT~\cite{hong2023metagpt} address how agents communicate, how tasks are divided into stages, and how standard procedures are embedded into workflows, respectively. The contribution here is complementary: while those frameworks define how agents coordinate, the focus here is on structuring the knowledge that agents depend on. A related but distinct approach is embedding-based retrieval over the codebase itself~\cite{tao2025rag}, as implemented in tools like Cursor's codebase indexing. These systems index \emph{code}; the codified context infrastructure indexes \emph{knowledge about code}---design intent, constraints, and failure modes not present in any single source file.

Benchmark efforts such as SWE-bench~\cite{jimenez2024swebench} and SWE-agent~\cite{yang2024sweagent} evaluate autonomous issue resolution on isolated tasks, while the Confucius Code Agent~\cite{wong2025confucius} addresses cross-session persistence through auto-generated notes that capture execution traces. The present work targets sustained, human-directed development, where specifications especially encode architectural intent and design constraints. The broader transition from code completion to LLM-assisted software engineering has prompted calls for formal scaffolding, grounding, and trust mechanisms~\cite{roychoudhury2025trust, hassan2025pillars}; the tiered architecture presented here offers one concrete realization of these principles.

\section{Architecture}
\label{sec:architecture}

\emph{Codified context infrastructure} is defined here as structured artifacts written explicitly for machine consumption---documents whose primary audience is an AI agent, not a developer. The architecture consists of three tiers, each with a distinct loading strategy and update frequency: \emph{hot memory} (the constitution, always loaded), \emph{domain specialists} (agents, invoked per task), and \emph{cold memory} (the knowledge base, retrieved on demand). Recent work has proposed treating prompts as engineered software artifacts~\cite{chen2025promptware}. The architecture presented here takes that idea further: rather than engineering individual prompts, it engineers project knowledge.

\begin{figure*}[t]
\centering
\begin{tikzpicture}[
    font=\sffamily,
    >=Stealth,
    tier/.style={
        draw=black!70,
        line width=0.8pt,
        rounded corners=4pt,
        minimum height=1.5cm,
        minimum width=3.8cm,
        text width=3.5cm,
        align=center,
        inner sep=5pt,
    },
    mechanism/.style={
        draw=black!60,
        line width=0.6pt,
        rounded corners=3pt,
        minimum height=0.9cm,
        align=center,
        inner sep=4pt,
        fill=white,
    },
    arrlabel/.style={
        font=\sffamily\scriptsize,
        text=black!65,
        inner sep=1pt,
    },
]

\def\colL{0}        
\def\colC{5.4}      
\def\colR{10.8}     


\node[
    draw=black!50,
    line width=0.6pt,
    rounded corners=8pt,
    fill=gray!5,
    minimum height=0.7cm,
    minimum width=2.0cm,
    font=\sffamily\small,
] (prompt) at (\colL, 0) {Human Prompt};

\node[
    draw=black!70,
    line width=0.9pt,
    rounded corners=4pt,
    fill=gray!8,
    minimum height=0.9cm,
    minimum width=3.5cm,
    font=\sffamily\small\bfseries,
] (session) at (\colC, 0) {Session};

\node[
    draw=black!50,
    line width=0.6pt,
    rounded corners=8pt,
    fill=gray!5,
    minimum height=0.7cm,
    minimum width=2.0cm,
    font=\sffamily\small,
] (result) at (\colR, 0) {Result};

\draw[->, line width=0.8pt, black!75]
    (prompt.east) -- (session.west);

\draw[->, line width=0.8pt, black!75]
    (session.east) -- (result.west);


\node[tier,
    fill=gray!30,
    draw=black!80,
    line width=1.4pt,
] (T1) at (\colL, -3.3) {
    {\small\bfseries Tier 1: Constitution}\\[2pt]
    {\footnotesize\textit{hot memory}}\\[1pt]
    {\scriptsize ${\sim}$660 lines}
};

\node[tier,
    fill=gray!15,
] (T2) at (\colC, -3.3) {
    {\small\bfseries Tier 2: Specialist Agent}\\[2pt]
    {\footnotesize\textit{domain experts}}\\[1pt]
    {\scriptsize 19 agents\;\textperiodcentered\;${\sim}$9,300 lines}
};

\node[tier,
    fill=gray!5,
] (T3) at (\colR, -3.3) {
    {\small\bfseries Tier 3: Knowledge Base}\\[2pt]
    {\footnotesize\textit{cold memory}}\\[1pt]
    {\scriptsize 34 docs\;\textperiodcentered\;${\sim}$16,250 lines}
};

\draw[->, line width=0.9pt, black!80, rounded corners=4pt]
    ($(session.south west) + (0.3, 0)$) -- ++(0, -0.55) -| (T1.north);
\node[font=\sffamily\scriptsize\itshape, text=black!70, anchor=south east]
    at ($(T1.north) + (-0.1, 0.15)$) {always loaded};

\node[mechanism,
    minimum width=3.2cm,
    text width=2.9cm,
] (mcp) at (8.1, -1.5) {
    {\scriptsize\bfseries MCP Retrieval}\\[-1pt]
    {\scriptsize keyword search over specs}
};

\coordinate (merge) at (\colC, -1.5);
\coordinate (t1mcp) at ($(T1.north) + (0.8, 0)$);
\draw[line width=0.6pt, dashed, black!50, rounded corners=4pt]
    (t1mcp) -- (t1mcp |- merge) -- (merge);
\draw[line width=0.6pt, dashed, black!50]
    (T2.north) -- (merge);
\draw[->, line width=0.6pt, dashed, black!50]
    (merge) -- (mcp.west);
\node[font=\sffamily\scriptsize\itshape, text=black!70, anchor=south]
    at ($(merge)!0.5!(mcp.west) + (-1.0, 0)$) {queried based on context};

\draw[->, line width=0.6pt, black!60, rounded corners=4pt]
    (mcp.east) -| (T3.north);
\node[font=\sffamily\scriptsize\itshape, text=black!65, anchor=south west]
    at ($(mcp.east) + (0.1, 0.05)$)
    {\texttt{find\_relevant\_context(task)}};

\draw[->, line width=0.6pt, black!60, rounded corners=4pt]
    (mcp.south) |- (T2.east);
\node[font=\sffamily\scriptsize\itshape, text=black!65, anchor=west]
    at ($(mcp.south) + (0.05, -0.25)$)
    {\texttt{suggest\_agent(task)}};


\node[mechanism,
    minimum width=3.2cm,
    text width=2.9cm,
] (trigger) at ($(T1.south)!0.5!(T2.south) + (0, -1.6)$) {
    {\scriptsize\bfseries Trigger Table}\\[-1pt]
    {\scriptsize file pattern $\rightarrow$ agent}
};

\draw[->, line width=0.6pt, black!60, rounded corners=4pt]
    (T1.south) -- ++(0, -0.4) coordinate (t1hook) -| ($(trigger.north west) + (0.15, 0)$);
\node[arrlabel, anchor=north]
    at ($(t1hook)!0.5!(t1hook -| trigger.north west) + (-0.45, -0.1)$) {\textit{consults before changes}};

\draw[->, line width=0.6pt, black!60, rounded corners=4pt]
    ($(trigger.north east) + (-0.15, 0)$) |- ($(T2.south) + (0, -0.4)$) -- (T2.south);

\end{tikzpicture}
\caption{Three-tier codified context infrastructure. A human prompt enters the session, which always loads the constitution (Tier~1, hot memory). Specialist agents (Tier~2) are invoked per task via the trigger table; the knowledge base (Tier~3, cold memory) is queried on demand through the MCP retrieval service. Decreasing fill intensity reflects decreasing load frequency.}
\label{fig:architecture}
\Description{Three-row architecture diagram. Top row: Human Prompt flows into a Session box, which flows to Result. Middle row: three tier boxes with differentiated loading arrows from Session. Tier 1 Constitution in darkest gray with heavy border, loaded via solid arrow labeled always loaded. Tier 2 Specialist Agent in medium gray, loaded via regular arrow labeled invoked per task. Tier 3 Knowledge Base in lightest gray, loaded via dashed arrow labeled on demand. Bottom row: Trigger Table mechanism between Tier 1 and Tier 2, MCP Retrieval Service mechanism between Tier 2 and Tier 3.}
\end{figure*}
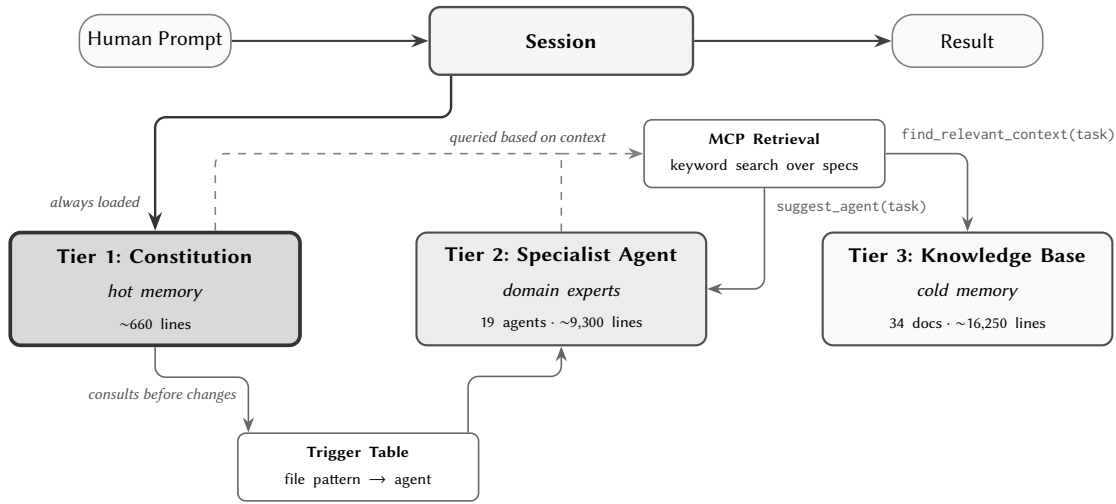

The tiers are not hermetically sealed. Specialized agents (Tier~2) embed substantial project-specific domain knowledge directly into their specifications---often constituting over half of agent content---rather than relying solely on retrieval from Tier~3. This intentional overlap emerged from the observation that agents operating in complex, bug-prone domains produced significantly more errors without pre-loaded context, consistent with the \emph{brevity bias} phenomenon~\cite{zhang2026ace}. Figure~\ref{fig:architecture} illustrates the relationships between tiers.

\subsection{Tier~1: Project Constitution (Hot Memory)}
\label{sec:constitution}

The constitution is a single Markdown file (${\sim}$660 lines) loaded automatically into every AI session. It defines code quality standards, naming conventions, build commands, architectural pattern summaries with references to detailed specifications in Tier~3, checklists for common operations, known failure modes, and orchestration protocols that route tasks to specialized agents.

The governing design constraint is conciseness: the constitution must fit entirely in every session without excessive context window consumption. Detailed subsystem documentation belongs in Tier~3, referenced by link. The constitution answers ``what rules must you always follow?''; Tier~3 answers ``how does subsystem X work in detail?''

\subsubsection{Orchestration Protocols}
\label{sec:orchestration}

The constitution embeds trigger tables that route tasks to the appropriate specialized agent (Tier~2) based on observable signals---primarily which files are being modified.

\begin{table}[tb]
\caption{Representative Orchestration Triggers}
\label{tab:triggers}
\footnotesize
\centering
\renewcommand{\arraystretch}{1.35}
\begin{tabular*}{\columnwidth}{@{\extracolsep{\fill}}llp{3.3cm}@{}}
\toprule
\textbf{Trigger} & \textbf{Signal} & \textbf{Agent} \\
\midrule
Pre-change & Network, sync & \texttt{network-protocol-\allowbreak designer} \\
Pre-change & Coordinates, camera & \texttt{coordinate-\allowbreak wizard} \\
Pre-change & Abilities end-to-end & \texttt{ability-designer} \\
Post-change & Architecture, design & \texttt{systems-designer} \\
Post-change & ECS or network files & \texttt{code-reviewer-\allowbreak game-dev} \\
\bottomrule
\end{tabular*}
\end{table}

Automatic routing removes the burden of the developer remembering which agent to invoke. The trigger table encodes institutional knowledge about which domain expertise each file area requires, addressing the inter-agent misalignment and planner-coder gap identified in prior work~\cite{cemri2025multiagent, lyu2025planner}. Routing compliance is reinforced by redundant encoding: the constitution requires the orchestrator to consult the trigger table before changes, and requires use of \texttt{suggest\_agent(task\_description)} via the MCP retrieval server when exploring unfamiliar code.

\subsection{Tier~2: Specialized Agents}
\label{sec:agents}

Nineteen agent specification files (Markdown, 115--1,233 lines each, ${\sim}$9,300 lines total) define domain-expert personas for specific areas of the codebase. The agents split into two capability classes: 8 higher-capability agents (${\sim}$5,700 lines, averaging ${\sim}$711 lines/agent) handling complex domains such as networking, architecture, and debugging, and 11 standard-capability agents (${\sim}$3,600 lines, averaging ${\sim}$327 lines/agent) for more focused tasks. Each specification declares a domain scope, available tools and permissions (some agents are read-only for safety), relevant Tier~3 documents, output format expectations, and common domain mistakes.

Agents function as domain-priming mechanisms: rich, structured context produces more reliable agent behavior~\cite{hong2023metagpt}. A networking agent reviewing damage synchronization code catches issues (missing client prediction, incorrect authority checks) that a general-purpose session would miss because the specification primes it with domain-specific failure modes.

\textbf{Knowledge embedding.} In the agents developed for this project, over half of each specification's content is project-domain knowledge (codebase facts, formulas, code patterns, known failure modes) rather than behavioral instructions (tools, permissions, output format). This creates intentional overlap with Tier~3, driven by three factors. Complex domains require complete mental models---the networking agent (915 lines, ${\sim}$65\% domain knowledge) embeds the full determinism theory because partial knowledge risks desynchronization bugs. Agents also accumulate symptom-cause-fix tables distilled from debugging sessions, codifying knowledge to prevent recurrence. Finally, some domains require a pre-synthesized view spanning multiple Tier~3 documents that would otherwise require 4--5 separate retrievals.

\textbf{Emergence pattern.} Agent creation was typically driven by observed failure patterns rather than upfront design. The first agents created were the network-protocol-designer and coordinate-wizard---the two domains with the highest debugging-session failure rates. When a category of task repeatedly required re-explaining the same domain knowledge, that knowledge was codified into an agent specification. The practical heuristic: if debugging a particular domain consumed an extended session without resolution, it was faster to create a specialized agent and restart the task than to continue the unguided session. An abbreviated agent specification is provided in Appendix~\ref{app:agent}.

\begin{table}[t]
\caption{Representative Agent Specifications and Creation Triggers}
\label{tab:agents}
\footnotesize
\centering
\renewcommand{\arraystretch}{1.35}
\begin{tabular*}{\columnwidth}{@{\extracolsep{\fill}}p{2.1cm}p{2.1cm}p{3.6cm}@{}}
\toprule
\textbf{Agent} & \textbf{Domain} & \textbf{Creation Trigger} \\
\midrule
\texttt{network-protocol-\allowbreak designer} & Sync, determinism & Recurring desync bugs needing full determinism re-explanation \\
\texttt{coordinate-\allowbreak wizard} & Isometric, camera & Persistent transform errors from mixed coordinate spaces \\
\texttt{code-reviewer-\allowbreak game-dev} & Post-change review & Regressions in ECS and networking after unreviewed changes \\
\texttt{level-designer} & Dungeon config, tiles & Brainstorming partner for procedural level design \\
\texttt{sprite-2d-artist} & Atlases, placeholders & Complex packing workflows across heterogeneous sprite formats \\
\bottomrule
\end{tabular*}
\end{table}

\subsection{Tier~3: Codified Context Base (Cold Memory)}
\label{sec:knowledge-base}

The knowledge base comprises 34 Markdown files (${\sim}$16,250 lines including the ${\sim}$1,600-line retrieval service), each documenting one subsystem. Three design decisions govern specification authorship: documents are written for AI consumption (explicit code patterns with file paths, parameter names, and expected behavior); specifications are living documents generated and updated by the AI at the developer's direction; and each document is scoped to a single subsystem to enable targeted retrieval.

\noindent\textbf{Specification format.} The following abbreviated example illustrates the structure of a knowledge base document:

\begin{lstlisting}[language={},basicstyle=\ttfamily\footnotesize]
# Deterministic RNG Synchronization

## Core Mechanism
CombatRng uses a deterministic hash:
  hash(seed, playerId, shotCounter) -> float [0,1)

## Correctness Pillars
| Pillar       | Requirement                       |
|--------------|-----------------------------------|
| Same seed    | Distributed at game start         |
| Same counter | Incremented per shot, synced      |
| Same time    | Use GetSyncedTime(), NOT local    |

## Known Failure Modes
| Symptom         | Cause               | Fix             |
|-----------------|----------------------|-----------------|
| Desync on crits | Using local time     | GetSyncedTime() |
| Inconsistent    | Counter not synced   | Sync shot count |
\end{lstlisting}

\noindent\emph{Full-length examples are available in the companion repository.}

\subsubsection{Knowledge Retrieval Service}
\label{sec:retrieval}

The knowledge base is served through a Model Context Protocol (MCP)~\cite{anthropic2025mcp} server (${\sim}$1,600 lines Python) that indexes specifications and provides five search tools:
\begin{itemize}[nosep,leftmargin=1em]
\item \texttt{list\_subsystems()}
\item \texttt{get\_files\_for\_subsystem(key)}
\item \texttt{find\_relevant\_context(task)}
\item \texttt{search\_context\_documents(query)}
\item \texttt{suggest\_agent(task)}
\end{itemize}
\noindent The current implementation uses keyword substring matching; semantic retrieval is discussed in Section~\ref{sec:future}.

\section{Evaluation}
\label{sec:evaluation}

\subsection{Methodology and Scope}
\label{sec:methodology}

This is a systems paper and experience report, where the primary contribution is the architecture rather than statistical evidence of effectiveness. The architectural description is complemented with: (a)~quantitative metrics on infrastructure scale and growth from the project's Git history; (b)~interaction metrics extracted from conversation history; (c)~four observational case studies; and (d)~qualitative observations on maintenance costs and failure modes. The case studies document instances where codified context was retrieved and applied across sessions with observable effects. No causal relationships are claimed; confounding factors including developer experience growth cannot be isolated.

\subsection{Scale and Growth}
\label{sec:scale}

The architecture was developed in the context of a real-time distributed system built across 283 sessions spanning 70 days of part-time development.

\begin{table}[t]
\caption{Project and Infrastructure Scale}
\label{tab:scale}
\small
\centering
\begin{tabular*}{\columnwidth}{@{\extracolsep{\fill}}lr@{}}
\toprule
\textbf{Metric} & \textbf{Value} \\
\midrule
Development period & 70 days (part-time) \\
Total commits & 148 \\
C\# source files & 405 \\
C\# lines of code & 108,256 \\
ECS systems & 45+ \\
ECS components & 55+ \\
Network message types & 35+ \\
Human prompts & 2,801 \\
Agent invocations & 1,197 \\
Agent turns & 16,522 \\
Development sessions & 283 \\
\midrule
\textbf{Context Infrastructure} & \textbf{Files / Lines / \% of Code} \\
\midrule
T1: Constitution & 1 / ${\sim}$660 / 0.6\% \\
T2: Specialized Agents & 19 / ${\sim}$9,300 / 8.6\% \\
T3: Knowledge Base & 34 / ${\sim}$16,250 / 15.0\% \\
\midrule
\textbf{Total context} & \textbf{54 / ${\sim}$26,200 / 24.2\%} \\
\bottomrule
\end{tabular*}
\end{table}

The knowledge-to-code ratio of 24.2\% reflects this project's complexity and domain, not a target or finding. A more actionable signal is agent behavior: when an agent produces inconsistent output or seems uncertain about a domain, the relevant specification is likely missing or stale.

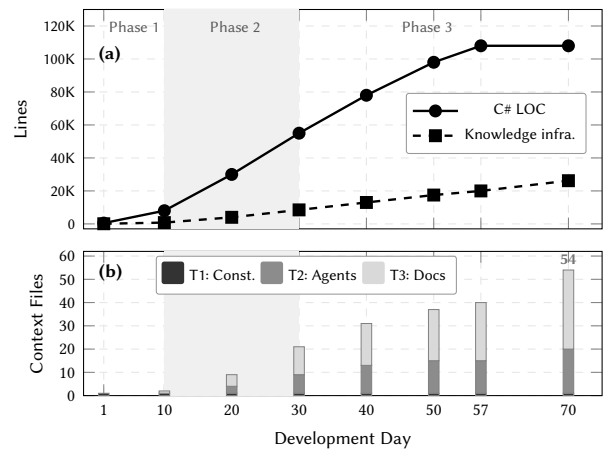
\begin{figure}[t]
\centering
\begin{tikzpicture}

\begin{axis}[
    name=topplot,
    font=\sffamily,
    width=\columnwidth,
    height=4.5cm,
    xmin=-2, xmax=75,
    ymin=-3000, ymax=130000,
    xtick={1,10,20,30,40,50,57,70},
    xticklabels={},  
    ytick={0,20000,40000,60000,80000,100000,120000},
    yticklabels={0,20K,40K,60K,80K,100K,120K},
    scaled y ticks=false,
    ylabel={Lines},
    grid=major,
    grid style={gray!20, thin, dashed},
    tick label style={font=\sffamily\scriptsize},
    label style={font=\sffamily\footnotesize},
    legend style={
        at={(0.97,0.62)},
        anchor=north east,
        font=\sffamily\scriptsize,
        draw=black!40,
        fill=white,
        rounded corners=1pt,
        inner sep=2pt,
    },
    clip=false,
    at={(0,0)},
    execute at begin axis={
        \fill[gray!12] (axis cs:10,-3000) rectangle (axis cs:30,130000);
    },
]

\node[font=\sffamily\scriptsize, text=black!65, anchor=north] at (axis cs:5.5,128000)
    {Phase 1};
\node[font=\sffamily\scriptsize, text=black!65, anchor=north] at (axis cs:20.5,128000)
    {Phase 2};
\node[font=\sffamily\scriptsize, text=black!65, anchor=north] at (axis cs:49,128000)
    {Phase 3};

\node[font=\sffamily\footnotesize\bfseries, anchor=north west] at (axis cs:-1,113000)
    {(a)};

\addplot[
    black,
    line width=0.9pt,
    mark=*,
    mark size=2pt,
] coordinates {
    (1, 500)
    (10, 8000)
    (20, 30000)
    (30, 55000)
    (40, 78000)
    (50, 98000)
    (57, 108000)
    (70, 108000)
};
\addlegendentry{C\# LOC}

\addplot[
    black,
    line width=0.9pt,
    dashed,
    mark=square*,
    mark size=2pt,
    mark options={solid},
] coordinates {
    (1, 100)
    (10, 800)
    (20, 4000)
    (30, 8500)
    (40, 13000)
    (50, 17500)
    (57, 20000)
    (70, 26200)
};
\addlegendentry{Knowledge infra.}

\end{axis}

\begin{axis}[
    font=\sffamily,
    width=\columnwidth,
    height=3.5cm,
    at={(topplot.south west)},
    anchor=north west,
    yshift=-0.3cm,
    xlabel={Development Day},
    ylabel={Context Files},
    xmin=-2, xmax=75,
    ymin=0, ymax=62,
    xtick={1,10,20,30,40,50,57,70},
    ytick={0,10,20,30,40,50,60},
    grid=major,
    grid style={gray!20, thin, dashed},
    tick label style={font=\sffamily\scriptsize},
    label style={font=\sffamily\footnotesize},
    ybar stacked,
    reverse stacked plots=false,
    bar width=4pt,
    legend style={
        at={(0.72,0.97)},
        anchor=north east,
        font=\sffamily\scriptsize,
        draw=black!40,
        fill=white,
        rounded corners=1pt,
        inner sep=2pt,
        legend columns=3,
    },
    legend image code/.code={
        \draw[#1, draw=none] (0cm,-0.1cm) rectangle (0.3cm,0.2cm);
    },
    clip=false,
    execute at begin axis={
        \fill[gray!12] (axis cs:10,0) rectangle (axis cs:30,62);
    },
]

\node[font=\sffamily\footnotesize\bfseries, anchor=north west] at (axis cs:-1,60)
    {(b)};

\addplot[
    fill=black!80,
    draw=black!50,
    line width=0.3pt,
] coordinates {
    (1, 1)
    (10, 1)
    (20, 1)
    (30, 1)
    (40, 1)
    (50, 1)
    (57, 1)
    (70, 1)
};
\addlegendentry{T1: Const.}

\addplot[
    fill=black!45,
    draw=black!50,
    line width=0.3pt,
] coordinates {
    (1, 0)
    (10, 0)
    (20, 3)
    (30, 8)
    (40, 12)
    (50, 14)
    (57, 14)
    (70, 19)
};
\addlegendentry{T2: Agents}

\addplot[
    fill=black!15,
    draw=black!50,
    line width=0.3pt,
] coordinates {
    (1, 0)
    (10, 1)
    (20, 5)
    (30, 12)
    (40, 18)
    (50, 22)
    (57, 25)
    (70, 34)
};
\addlegendentry{T3: Docs}

\node[font=\sffamily\scriptsize\bfseries, text=black!55, anchor=south] at (axis cs:70,53)
    {54};

\end{axis}

\end{tikzpicture}
\caption{Infrastructure growth across eight milestones reconstructed from Git history. (a)~C\# code (solid) and knowledge infrastructure (dashed) lines over time, with three development phases as alternating background bands. (b)~Context file counts by tier: constitution (T1, darkest), specialist agents (T2, medium), and knowledge base documents (T3, lightest).}
\label{fig:growth}
\Description{Two vertically stacked panels sharing an x-axis of Development Day with eight milestones. Panel (a) shows two lines: solid black for C-sharp lines of code rising from 500 to 108,000, and dashed black for knowledge infrastructure lines rising from 100 to 26,200. Three phase regions shown as alternating white and light gray vertical bands. Panel (b) shows stacked bar chart of context file counts by tier: T1 Constitution as a constant single-file segment at the bottom of each bar, T2 Agents growing from 0 to 19, and T3 Knowledge Base documents growing from 0 to 34, reaching a total of 54 files.}
\end{figure}

This final-state snapshot does not convey how the infrastructure evolved. Figure~\ref{fig:growth} reconstructs the trajectory from Git history across three phases: Phase~1 (days~1--10) used only a ${\sim}$100-line constitution; Phase~2 (days~11--30) saw the emergence of specifications and initial agents for high-failure-rate domains; Phase~3 (days~31--57) integrated the MCP retrieval service and expanded the agent pool as new domains required specialization. Raw milestone data is available in the companion repository.

\subsection{Interaction Metrics}
\label{sec:interaction}

A \emph{session} is defined as a single conversation instance with the agentic coding tool; the 283 sessions span 70 days and consist primarily of development work, with a smaller proportion devoted to brainstorming, debugging, and other miscellaneous tasks. Interaction data was extracted from 1,457 JSONL conversation history files (a representative snapshot; some early files were lost during a cache cleanup, and agent chain data is available only for a 31-day window within the measurement period). The extraction methodology, scripts, and dataset schema are described in the companion repository.

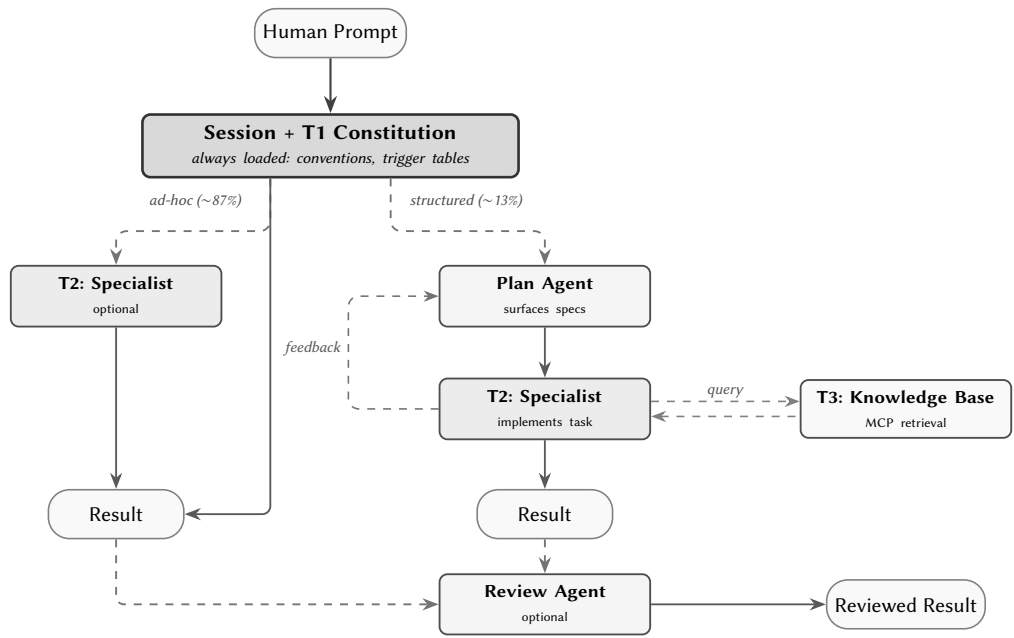
\begin{figure*}[t]
\centering
\begin{tikzpicture}[
    font=\sffamily,
    >=Stealth,
    box/.style={
        draw=black!70,
        line width=0.7pt,
        rounded corners=3pt,
        minimum height=0.75cm,
        align=center,
        inner sep=4pt,
        font=\sffamily\small,
    },
    terminal/.style={
        draw=black!50,
        line width=0.6pt,
        rounded corners=8pt,
        fill=gray!5,
        minimum height=0.65cm,
        minimum width=1.8cm,
        font=\sffamily\small,
    },
    arrlabel/.style={
        font=\sffamily\scriptsize,
        text=black!65,
        inner sep=1pt,
    },
    pctlabel/.style={
        font=\sffamily\scriptsize\itshape,
        text=black!70,
    },
]

\def\colL{-2.5}    
\def\colR{3.2}     
\def\colKB{8.0}    

\node[terminal] (prompt) at (0.35, 0) {Human Prompt};

\node[box, fill=gray!30, line width=1.0pt, draw=black!80,
    minimum width=5.0cm, text width=4.7cm] (session) at (0.35, -1.5)
    {{\bfseries Session + T1 Constitution}\\[-1pt]
     {\scriptsize\textit{always loaded: conventions, trigger tables}}};

\draw[->, line width=0.8pt, black!75]
    (prompt.south) -- (session.north);


\node[pctlabel] at ($(session.south) + (-1.8, -0.3)$)
    {ad-hoc (${\sim}$87\%)};

\node[box, fill=gray!15,
    minimum width=2.8cm, text width=2.5cm] (adhoc-agent) at (\colL, -3.5)
    {{\footnotesize\bfseries T2: Specialist}\\[-1pt]
     {\tiny optional}};

\draw[->, line width=0.7pt, dashed, black!55, rounded corners=4pt]
    ($(session.south) + (-0.8, 0)$) -- ++(0, -0.7) -| (adhoc-agent.north);

\node[terminal] (result-L) at (\colL, -6.4) {Result};

\draw[->, line width=0.7pt, black!65]
    (adhoc-agent.south) -- (result-L.north);

\draw[->, line width=0.7pt, black!65, rounded corners=3pt]
    ($(session.south) + (-0.8, 0)$) |- (result-L.east);

\node[pctlabel] at ($(session.south) + (1.8, -0.3)$)
    {structured (${\sim}$13\%)};

\node[box, fill=gray!8,
    minimum width=2.8cm, text width=2.5cm] (plan) at (\colR, -3.5)
    {{\footnotesize\bfseries Plan Agent}\\[-1pt]
     {\tiny surfaces specs}};

\draw[->, line width=0.7pt, dashed, black!55, rounded corners=4pt]
    ($(session.south) + (0.8, 0)$) -- ++(0, -0.7) -| (plan.north);

\node[box, fill=gray!15,
    minimum width=2.8cm, text width=2.5cm] (specialist) at (\colR, -5.0)
    {{\footnotesize\bfseries T2: Specialist}\\[-1pt]
     {\tiny implements task}};

\draw[->, line width=0.7pt, black!65]
    (plan.south) -- (specialist.north);

\node[box, fill=gray!5,
    minimum width=2.8cm, text width=2.5cm] (kb) at (\colKB, -5.0)
    {{\footnotesize\bfseries T3: Knowledge Base}\\[-1pt]
     {\tiny MCP retrieval}};

\draw[->, line width=0.6pt, dashed, black!50]
    ($(specialist.east) + (0, 0.1)$) -- ($(kb.west) + (0, 0.1)$)
    node[arrlabel, above, midway] {\textit{query}};
\draw[<-, line width=0.6pt, dashed, black!50]
    ($(specialist.east) + (0, -0.1)$) -- ($(kb.west) + (0, -0.1)$);

\draw[->, line width=0.6pt, dashed, black!55, rounded corners=4pt]
    (specialist.west) -- ++(-1.2, 0) |- (plan.west)
    node[arrlabel, anchor=east, pos=0.5, xshift=-0.05cm, yshift=-0.7cm] {\textit{feedback}};

\node[terminal] (result-R) at (\colR, -6.4) {Result};

\draw[->, line width=0.7pt, black!65]
    (specialist.south) -- (result-R.north);

\node[box, fill=gray!8,
    minimum width=2.8cm, text width=2.5cm] (review) at (\colR, -7.6)
    {{\footnotesize\bfseries Review Agent}\\[-1pt]
     {\tiny optional}};

\draw[->, line width=0.7pt, dashed, black!55]
    (result-R.south) -- (review.north);

\draw[->, line width=0.7pt, dashed, black!55, rounded corners=3pt]
    (result-L.south) -- (\colL, -7.6) -- (review.west);

\node[terminal] (final-result) at (\colKB, -7.6) {Reviewed Result};

\draw[->, line width=0.7pt, black!65]
    (review.east) -- (final-result.west);

\end{tikzpicture}
\caption{Typical session workflows for development tasks. A prompt enters the session with the always-loaded constitution (Tier~1). Ad-hoc sessions (${\sim}$87\%) produce a result directly or optionally invoke a specialist agent; structured sessions (${\sim}$13\%) follow a modified plan-execute-review cycle with specialist agents (Tier~2) and knowledge base retrieval (Tier~3). Knowledge base retrieval is available in both modes but shown only on the structured path. Dashed arrows indicate optional steps.}
\label{fig:workflow}
\Description{Top-to-bottom knowledge flow diagram showing two parallel paths from a Session plus T1 Constitution box. Left path: ad-hoc sessions at 87 percent, with a dashed arrow to an optional T2 Specialist Agent, then to Result. Right path: structured sessions at 13 percent, flowing through Plan Agent to T2 Specialist Agent with bidirectional dashed arrows to T3 Knowledge Base, then optionally to Review Agent which feeds back to Plan Agent, ending at Result. Solid arrows show always-present steps, dashed arrows show optional steps.}
\end{figure*}

The interaction dataset comprises 2,801 human prompts, 1,197 agent invocations, and 16,522 autonomous agent turns across these sessions (19,323 total interactions; ${\sim}$9.9 human prompts per session). Figure~\ref{fig:workflow} illustrates the typical session workflows and retrieval patterns.

The orchestration protocols supported two modes: \emph{structured sessions} (${\sim}$13\%) following a plan-execute-review workflow (Figure~\ref{fig:workflow}), and \emph{ad-hoc sessions} (${\sim}$87\%) involving direct implementation or debugging. Each human prompt produced approximately 6 autonomous agent turns through agent-to-agent chaining---comprising file exploration, tool calls, plan review, and subject-matter agent consultations. Of 757 classifiable agent invocations, 432 (57\%) were project-specific specialists defined in the context infrastructure and 325 were built-in tool agents. The most frequently invoked specialists were the code reviewer (154 invocations) and the network-protocol-designer (85 invocations), indicating that the primary uses of orchestration were quality gating and networking correctness.

Over 80\% of human prompts were 100 words or fewer, consistent with the hypothesis that pre-loaded context reduces the need for in-prompt explanation. Meta-infrastructure prompts---building the knowledge architecture itself---accounted for 4.3\% of substantive prompts, representing the direct overhead of the approach.

\subsection{Case Studies}
\label{sec:case-studies}

The following four case studies illustrate distinct roles that codified context plays in development outcomes. Cases were selected for qualitative diversity, each illustrating a distinct mechanism (coordination, experience capture, gap detection, domain-expert diagnosis). The knowledge base was actively used throughout development: 1,478 MCP retrieval calls across 218 sessions, with 194 agent conversations explicitly reading knowledge base documents.

\subsubsection{Case Study~1: Save System---Codified Context as Coordination Document}
\label{sec:casestudy1}

The project's save system uses a two-tier architecture: a disk tier for permanent player data and a memory tier for temporary state during level transitions. Writing to the wrong tier causes subtle data corruption---temporary buffs persisting permanently, or gold rewards lost on restart. The \texttt{save-system.md} specification (283 lines, Tier~3) documents the two-tier architecture, autosave trigger points, the gold checkpoint/rollback mechanism, and data flow between game states.

This was the most-referenced specification in the project, appearing in 74 sessions and 12 agent conversations. Five subsequent features touching persistence---including an item system overhaul, a token upgrade system, and shop refactoring---were implemented by agents with access to this specification, and the two-tier pattern was consistently applied correctly. Coordination across 74 independent sessions over four weeks produced no save-related bugs.

\subsubsection{Case Study~2: UI Sync Routing---Codified Context as Captured Experience}
\label{sec:casestudy2}

The shop synchronization system, implemented before any networking UI specification existed, applied unreliable delivery uniformly---reasonable for high-frequency game state, but incorrect for UI state machines. Shop open/close events were silently dropped under packet loss, leaving clients with phantom overlays. Switching to reliable delivery for all messages fixed the drops but introduced excessive bandwidth for cursor position updates sent at 60~Hz. The developer manually diagnosed and corrected delivery mode assignments through multiple iterations.

After the shop implementation stabilized, the lessons were captured in \texttt{ui-sync-patterns.md} (126 lines), documenting three routing topologies, a delivery mode decision tree, and a dual delivery pattern. The specification was subsequently referenced in approximately 10 sessions and 25+ agent conversations. The next networked UI feature (a radial selection dial) correctly applied the dual delivery pattern on the first implementation attempt, following the specification's decision tree directly. The specification prevented the AI from re-deriving through trial-and-error what the shop implementation had already established.

\subsubsection{Case Study~3: Drop System---Knowledge Gap Detection}
\label{sec:casestudy3}

When refactoring the equipment system from a single item type to multiple composable types, a search for drop system documentation via the retrieval service returned zero results---no specification existed. The null result was itself informative: an entire subsystem had been built without ever being documented, and the specification needed to be created before the refactor could proceed safely.

Rather than proceeding to code changes, the session first created \texttt{drop-system.md} by reading approximately ten source files to document the existing architecture, revealing legacy code that would need to be addressed during the refactor. The developer then used the specification as a design document, proposing changes against the now-documented current state and having two specialist agents review the proposal. The refactor touched 14 source files, after which the specification was updated and subsequently referenced in dozens of later sessions. The upfront cost of documenting the subsystem---a single session---was repaid by the velocity of every subsequent interaction: agents working on loot, equipment, or inventory features could retrieve the drop system's architecture on demand rather than re-reading source files or receiving manual explanation. Maintaining a robust knowledge base converts a one-time documentation effort into a persistent acceleration of all downstream work.

\subsubsection{Case Study~4: Deterministic RNG---Agent Domain Knowledge in Collaborative Debugging}
\label{sec:casestudy4}

The project uses a deterministic random number generator (\texttt{CombatRng}) to ensure that host and client compute identical outcomes for the same game event. If any input to the hash function diverges between machines, players see different results---a desynchronization bug that is difficult to diagnose because each machine's output appears locally correct.

A time synchronization refactor, replacing local timestamps with network-reconciled time across 12 source files, introduced a determinism failure. The debugging session spanned five context window exhaustions and 84 code edits, with the bug peeling back in stages. The \texttt{network-protocol-designer} agent was invoked at inflection points after the developer had narrowed the problem space through debug log analysis. The agent's specification (${\sim}$915 lines) embeds the project's determinism theory: correctness pillars, hash function constraints, and known numerical edge cases. Drawing on this embedded knowledge, the agent identified three issues that had eluded five prior attempts: a guard clause that silently failed under normal conditions, two internal clocks updating at different rates, and a sign error in the clock correction formula. The agent concluded that using time as an input to the combat RNG hash was fundamentally counterproductive---small timing discrepancies between host and client would always risk divergent hash outputs. Replacing the time-bucket parameter with a synchronized shot counter resolved the bug, a recommendation that required understanding of hash sensitivity to timing granularity pre-loaded in the specification rather than derived during the session.

\section{Discussion}
\label{sec:discussion}

Across the development period, AI automation consistently eliminated tedium (implementation, rendering, wiring) but not judgment (design decisions, aesthetic evaluation, architecture). Single-file manifests support this division of labor early on, but as a project grows in complexity, agents lose coherence and the developer is increasingly pulled back into resolving routine implementation errors that the agent should be handling. The context infrastructure preserves this division at scale: persistent, machine-readable specifications keep agents producing correct, convention-adherent code even as the codebase grows, freeing the developer to remain focused on design and judgment. Experienced developers would likely derive different value from the architecture: less in preventing basic mistakes, more in maintaining consistency across a codebase too large for any single person to hold in working memory. Well-maintained project knowledge compounds: each documented subsystem accelerates not only its own future modifications but every adjacent feature that depends on it.

\subsection{Guidelines for Practitioners}
\label{sec:guidelines}

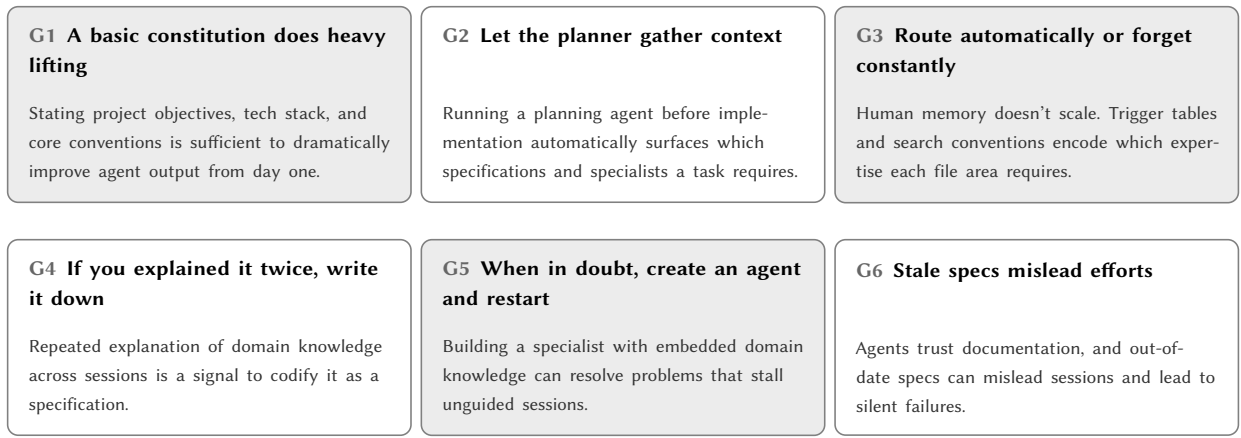
\begin{figure*}[t]
\centering
\begin{tikzpicture}[
    font=\sffamily,
    card/.style={
        draw=black!50,
        line width=0.6pt,
        rounded corners=4pt,
        minimum height=2.6cm,
        text width=4.8cm,
        align=left,
        inner sep=8pt,
        anchor=north west,
    },
    oddcard/.style={card, fill=gray!15},
    evencard/.style={card, fill=white},
    cardnum/.style={
        font=\sffamily\footnotesize\bfseries,
        text=black!60,
    },
]

\def\colA{0}
\def\colB{5.5}
\def\colC{11.0}
\def\rowA{0}
\def\rowB{-3.1}

\node[oddcard] (c1) at (\colA, \rowA) {%
    \parbox[t][0.9cm][t]{4.8cm}{{\small\bfseries\color{black!60} G1}\hspace{4pt}{\small\bfseries A basic constitution does heavy lifting}}\par\vspace{3pt}
    {\footnotesize\color{black!80} Stating project objectives, tech stack, and core conventions is sufficient to dramatically improve agent output from day one.}
};

\node[evencard] (c2) at (\colB, \rowA) {%
    \parbox[t][0.9cm][t]{4.8cm}{{\small\bfseries\color{black!60} G2}\hspace{4pt}{\small\bfseries Let the planner gather context}}\par\vspace{3pt}
    {\footnotesize\color{black!80} Running a planning agent before implementation automatically surfaces which specifications and specialists a task requires.}
};

\node[oddcard] (c3) at (\colC, \rowA) {%
    \parbox[t][0.9cm][t]{4.8cm}{{\small\bfseries\color{black!60} G3}\hspace{4pt}{\small\bfseries Route automatically or forget constantly}}\par\vspace{3pt}
    {\footnotesize\color{black!80} Human memory doesn't scale. Trigger tables and search conventions encode which expertise each file area requires.}
};

\node[evencard] (c4) at (\colA, \rowB) {%
    \parbox[t][0.9cm][t]{4.8cm}{{\small\bfseries\color{black!60} G4}\hspace{4pt}{\small\bfseries If you explained it twice, write it down}}\par\vspace{3pt}
    {\footnotesize\color{black!80} Repeated explanation of domain knowledge across sessions is a signal to codify it as a specification.}
};

\node[oddcard] (c5) at (\colB, \rowB) {%
    \parbox[t][0.9cm][t]{4.8cm}{{\small\bfseries\color{black!60} G5}\hspace{4pt}{\small\bfseries When in doubt, create an agent and restart}}\par\vspace{3pt}
    {\footnotesize\color{black!80} Building a specialist with embedded domain knowledge can resolve problems that stall unguided sessions.}
};

\node[evencard] (c6) at (\colC, \rowB) {%
    \parbox[t][0.9cm][t]{4.8cm}{{\small\bfseries\color{black!60} G6}\hspace{4pt}{\small\bfseries Stale specs mislead efforts}}\par\vspace{3pt}
    {\footnotesize\color{black!80} Agents trust documentation, and out-of-date specs can mislead sessions and lead to silent failures.}
};

\end{tikzpicture}
\captionsetup{singlelinecheck=false}
\caption{Practitioner guidelines for using a codified context infrastructure.}
\label{fig:findings}
\Description{A 2-row by 3-column grid of six equal-sized rounded rectangles spanning the full text width. Cards are numbered G1 through G6 and alternate between light gray and white fill. G1: A basic constitution does heavy lifting. G2: Let the planner gather context. G3: Route automatically or forget constantly. G4: If you explained it twice write it down. G5: When in doubt create an agent and restart. G6: Stale specs mislead efforts.}
\end{figure*}

Figure~\ref{fig:findings} summarizes key practical guidelines distilled from this project. For example: start the constitution early---even a minimal file stating project objectives, tech stack, and core conventions prevents an entire class of AI mistakes from day one; and write specifications for the agent, not for humans, with file paths, function names, and explicit ``do this / don't do this'' instructions. Two further guidelines address maintenance: treat specifications as load-bearing, because agents trust documentation absolutely and out-of-date specs cause silent failures; and monitor agent confusion as a diagnostic signal that the relevant specification is missing or stale. Factory agents for bootstrapping each tier are provided in the companion repository.

\subsection{Maintenance Cost}
\label{sec:maintenance}

In practice, specification updates were performed in the same session as code changes---typically one or two prompts directing the AI to update the relevant document, adding roughly 5 minutes per session when a specification was affected. This per-session overhead was supplemented by a biweekly review pass across all context documents, each taking approximately 30--45 minutes. Total maintenance overhead averaged approximately 1--2 hours per week.

Specification staleness was the primary failure mode. When a subsystem's implementation changes, its specification must be updated or the AI will generate code based on stale information. On at least two occasions, outdated context documents caused agents to generate code that conflicted with recent refactors. In one instance, a combat specification referenced legacy stat fields that had been migrated to a computed stats system, causing the agent to wire damage calculations through a deprecated path. Both issues were caught during the same session but required manual diagnosis---the agent's output appeared syntactically correct, and the errors only surfaced during testing.

A context drift detector (Python, session-start hook), included in the companion repository, partially automates this process by parsing recent Git commits against the retrieval service's subsystem-to-file mapping and injecting a warning into session context when source files change without corresponding specification updates.

\subsection{Threats to Validity and Future Work}
\label{sec:future}

\textbf{Single-developer, single-project evaluation.} The architecture was developed by a single developer on a single project; its effectiveness in team settings, other project types, or larger scales has not been evaluated. The architecture may benefit team environments: agents and orchestrators informed by up-to-date specifications would be aware of recent changes across the codebase without requiring explicit communication between team members. It is worth noting that the project domain (real-time distributed simulation) demands more extensive documentation than many application types, which may limit generalizability to simpler projects.

\textbf{Observational methodology.} The evaluation relies on observational case studies, not controlled experiments. It is not possible to quantify with statistical rigor how much the architecture improved development speed or code quality. The before/after comparisons in Case Studies~2 and~3 reflect observed changes over time, not randomized trials.

\textbf{Tool-specific implementation.} The implementation uses Claude Code with MCP support. The architectural \emph{principles} (tiered knowledge organization, hot/cold separation, domain-specialist routing) apply to any agentic coding tool that supports session-start configuration and on-demand retrieval, but the specific implementation is tied to one tool's ecosystem. Transferability has not been evaluated.

\textbf{Future directions.} Controlled benchmarking---measuring task completion rates and error rates with and without each architecture tier---is the most immediate priority. The drift detector could be extended with semantic diff analysis to detect when specification content contradicts changed code. Replacing keyword matching with embedding-based retrieval~\cite{tao2025rag} would improve precision at scale. Multi-project and team-scale evaluations would assess transferability and whether shared context infrastructure reduces onboarding time. Longitudinal study of how the knowledge-to-code ratio evolves would clarify whether it stabilizes, grows, or requires periodic pruning. Factory agents for bootstrapping the architecture on new projects are included in the companion repository.

\section{Conclusion}
\label{sec:conclusion}

The core insight of this work is that structured access to project-specific knowledge can substantially improve the consistency of AI-generated code, and that this knowledge can be organized into distinct tiers with different loading strategies and update frequencies. The tiered architecture presented here---a hot-memory constitution, specialized domain-expert agents, and a cold-memory knowledge base---treats project documentation as infrastructure rather than artifact: living specifications that AI agents depend on to produce correct, convention-adherent code. This architecture supported a single developer in constructing a 108,000-line distributed system in under 70 days of part-time development using AI agents as the sole code-generation tool.

The case studies illustrate four distinct mechanisms by which codified context improves development outcomes: specifications as inter-session coordination documents enabling 74 sessions of consistent persistence behavior (Case Study~1), captured experience preventing repeated trial-and-error across 10+ subsequent sessions (Case Study~2), documentation as an investment that converts one-time effort into persistent development velocity (Case Study~3), and embedded domain knowledge enabling collaborative debugging of subtle cross-cutting bugs (Case Study~4).

The context infrastructure itself can be AI-generated under human architectural direction---the human's role is designing the knowledge structure and deciding what to capture. As AI coding agents become more capable, this architecture is particularly relevant to domain experts building software beyond their primary expertise, where codified context compensates for gaps in engineering experience. The author is currently applying the framework to a drug discovery project as an initial test of cross-domain transferability.

The companion framework repository is available at \url{https://github.com/arisvas4/codified-context-infrastructure}, including representative agent specifications, the MCP retrieval server, example constitution and knowledge base documents, three factory agents for bootstrapping the architecture in new projects, and all prompt classification scripts referenced in this paper.

\begin{acks}
This work was conducted independently with no external funding. An initial draft was prepared with AI assistance and subsequently revised, restructured, and verified by the author.
\end{acks}

\bibliographystyle{ACM-Reference-Format}
\bibliography{references}

\appendix

\section{Tier Statistics}
\label{app:stats}

The agent specialization table is provided in Section~\ref{sec:agents}. The following table summarizes the five largest knowledge base documents (Tier~3).

\begin{table}[h!]
\caption{Knowledge Base Documents (Tier~3)---Top 5 by Size}
\label{tab:kb-top5}
\small
\centering
\begin{tabular*}{\columnwidth}{@{\extracolsep{\fill}}llr@{}}
\toprule
\textbf{Document} & \textbf{Subsystem} & \textbf{Lines} \\
\midrule
\texttt{dungeon-generation.md} & Procedural generation & ${\sim}$1,286 \\
\texttt{hud-blueprint.md} & HUD layout & ${\sim}$1,134 \\
\texttt{enemy-combat-system.md} & Enemy attacks & ${\sim}$779 \\
\texttt{boss-fight-framework.md} & Boss encounters & ${\sim}$722 \\
\texttt{architecture.md} & Core architecture & ${\sim}$690 \\
\bottomrule
\end{tabular*}
\end{table}

\section{Example Agent Specification}
\label{app:agent}

The following is an abbreviated specification for the \texttt{coordinate\mbox{-}\allowbreak wizard} agent, a read-only diagnostic agent for coordinate transform debugging.

\begin{lstlisting}[language={},escapechar=|]
---
name: coordinate-wizard
description: Isometric coordinate and camera
             transform specialist.
tools: Read, Grep, Glob, Bash,
       mcp__context7__get_files_for_subsystem,
       mcp__context7__search_context_documents
model: higher-capability
---

**This agent is READ-ONLY for diagnostics.**

# COORDINATE SPACES

| Space      | Description              | Use            |
|------------|--------------------------|----------------|
| World      | Rect grid (0,0=top-left) | Positions      |
| Grid       | Integer tile indices     | Tilemap        |
| Isometric  | 2:1 dimetric projection  | Rendering      |
| Virtual    | 1920x1080 fixed          | Camera, UI     |
| Physical   | Actual display pixels    | Final render   |

## COMMON BUG PATTERNS

| Symptom            | Cause               | Fix            |
|--------------------|----------------------|----------------|
| Wrong position     | No IsoCorrection     | Add correction |
| Wrong click target | No inverse transform | Use ScreenTo.. |
| Resolution offset  | Physical screen ctr  | VirtualToPhys  |
\end{lstlisting}

\emph{Abbreviated from the full 328-line specification. The complete version includes rendering patterns, conversion formulas, mouse picking chains, depth sorting, and debugging workflows. See the companion repository for other full examples.}

\end{document}